\documentclass[aps,secnumarabic,showpacs,nobibnotes,showkeys,superscriptaddress,
amssymb,prd,epsfigs,amsfonts,nofootinbib,amsmath,apsfonts,twoculumn]{revtex4}
\usepackage{graphicx}
\usepackage{dcolumn}
\usepackage{bm}
\begin{document}
\title{\Large\bf Solutions of the Renormalisation Group Equation 
in Minimal Supersymmetric Standard Model(MSSM).}
\author{B. B. Deo}
\email{bdeo@iopb.res.in}
\affiliation{Department of Physics , Utkal University, Bhubaneswar-751004, India.}
\author{L. Maharana}
\email{lmaharan@iopb.res.in}
\affiliation{Department of Physics , Utkal University, Bhubaneswar-751004, India.}
\author{P. K. Mishra}
\affiliation{Department of Physics, S.C.S. College, Puri-752001,  India}
\begin{abstract}
Renormalisation Group Equation(RGE) for color and top couplings sector of 
MSSM has been solved. The mass of the top comes out to be 
180.363$\pm$ 10.876 GeV and $\beta_{top}\cong \frac{\pi}{4}$.
It is conjectured that the masses of the other 11 fermions and the 
CKM phase angle $\phi$ can be theoretically estimated. The results 
confirm the fact that the quarks and leptons have been created having equal mass
$\sim$ 115 GeV at the MSSM GUT scale $\sim ~2.2\times 10^{16}$ GeV.
\end{abstract}
\maketitle
The Standard Model for Electroweak and Strong interaction has been proposed
with many variations like Standard Model(SM), two Higgs Doublet Model(2HDM),
Minimal Supersymmetric Standard Model(MSSM) and such others~\cite{r1a}. 
Most of them have been phenomenologically successful but this success 
has brought in relatively large number of free parameters~\cite{r1}. 
Besides the gauge couplings of $SU_C(3)$,
$SU_L(2)$ and $U_Y(1)$, in the matter sector there are Higgs and Yukawa couplings~
\cite{r1,r2}. The aim of a good model building is not only to reduce the number 
of free parameters but should  perturbatively tenable and obtain result in 
the agreement with experiment. To have an idea of the parameters, the 
vertex function of the model satisfy the renormalisation group equation~\cite{r3},
\begin{equation}
\left [
\mu\frac{\partial~~}{\partial{\mu}}
+\beta_{g_i}\frac{\partial~~~}{\partial{\beta_{g_i}}}
+\beta_{G_i}\frac{\partial~~~}{\partial{\beta_{G_i}}}-\sum_a\gamma_a N_a
\right ]\Gamma=0.\label{e1}
\end{equation}
The notations are usually taken from reference~\cite{r3}. To give an example of what
we imply by reliable calculation is to note that the gauge coupling coefficients
$\beta_{g_i}$ satisfy the following equations in MSSM which we shall consider in this
letter,
\begin{equation}
16\pi^2\frac{dg_i(t)}{dt~~~~}~=~c_ig_i^3(t),
~~~~\text{with}~~ t=log(\mu/M_Z),~~~~~~~i=1,2,3.\label{e2}
\end{equation}
where $\mu$ is the mass scale of the theory.
$c_1$=6.6, and $c_2$=1. This means that the $SU_L(2)$ and U(1)$_Y$ groups are 
not asymptotically free but $c_3$=$-$3 means that the color  $SU_C(3)$ 
is asymptotically free. Therefore one can make a perturbative expansion in 
powers of $g_3$.

It is not certain that the other $\beta$-coefficients obtained from product 
groups are such that the theory is asymptotically free and one can find 
reliable perturbative results. There is sufficient evidence to show that 
the top quark mass coupling decreases as energy increases like $g_3$. So from 
the entire parameter space of SM, we isolate a 
small region containing the top coupling and $g_3$ of the color group. The 
strategy is to see that if the experimentally acceptable results are obtained 
in this small region; we can then continue this region analytically to gradually 
encroach and cover the entire parameter space so that the perturbative result 
will be meaningful everywhere. This was also the basis of reduction of couplings
technique of Zimmermann et al~\cite{r3}. They developed a technique
of reduction of coupling with one Higgs doublet and  n generations and have 
shown that the cancellation of divergeces in the Higgs propagator is compatible
with renormalisation group invariance and reduction of couplings. For three
generations, they have predicted $m_{top}$=81 GeV and $m_H$=61 GeV with 
estimated error about 10-15\%.

Pendleton and Ross~\cite{r4a} have exclusively extended the Kube, Siebold and 
Zimmermann's~\cite{r4b} work to include the other gauge coupling parameters
$g_1$, $g_2$ and $g_3$ in the same type of standard model. Their result, 
taking $g_3$ coupling alone, is also~~ $m_{top}=\sqrt{\frac{2}{9}}g_3~v$=81 GeV.
Theoretical prediction of the mass of the top, before it was discovered,
have been made by several authors~\cite{r3a}. Faraggi has obtained 
$m_{top}\simeq 175 - 180$ GeV in Superstring derived standard like model, 
but, in getting this result, he has taken $\frac{m_{bottom}}{m_{top}}=\frac{1}{8}$
at the unification scale. We attempt here to extend earlier works to 
include the successful MSS Model into Yukawa coupling parameters calculations.

It will be helpful to state the gauge sector values ~\cite{r5} for later use.
We shall take $t_X$=log(M$_X$/M$_Z$)=33.0, $4\pi$/g$^2_U$=24.6 and $M_S=M_Z$.
The values of $g_1^2,~g_2^2,~g_3^2$ at $m_Z$ obtained from the R.G. equation  
above, are $4\pi/g_1^2=59.24,~4\pi/g_2^2=29.85$ and ~$4\pi/g_3^2$=8.85.
These are reasonable and consistent with the experimental results.

In this letter, we are primarily concerned in the top mass within the top-color
sector. In the renormalisation group equation for the top in MSSM, 
there is an added complication of having two Higgs. Therefore we write
for the top~\cite{r4}
\begin{equation}
m_{top}=M_{top}(t)~v_{top}(t).\label{e3}
\end{equation}
Conventionally,
\begin{equation}
v_{top}(t)= v~\text{sin}(\beta_{top}(t)),\label{e4}
\end{equation}
where $v$=174 GeV. 

Letting $g_1=g_2=0$, $m_{lepton}=0$, $m_{quark}$=0, except the top,
the MSSM, RGE is
\begin{eqnarray}
16\pi^2\frac{d m_{top}}{dt~~~}&=&\left [ -\frac{16}{3}g_3^2 + 
3M^2_{top}\right ] m_{top},\nonumber\\
&=&\left [ -\frac{16}{3}g_3^2 +
3\frac{m_{top}^2}{v_{top}^2}\right ] m_{top},\label{e5}\\
16\pi^2\frac{d g_3}{dt~}&=&-3g_3^3,\label{e6}\\
\text{and}~~~~~~~16\pi^2\frac{d v_{top}}{dt~~~}&=&\left [\frac{3}{20}g^2_1 
+\frac{3}{4}g^2_2 +0.0\times g^2_3
-3M_{top}^2\right ] v_{top},\\
&=&-3M_{top}^2~v_{top}. \label{e7}
\end{eqnarray}

The top Yukawa and color sector equations, which have to be solved 
self consistently, become 
\begin{eqnarray}
g_3^3\frac{dm_{top}}{dg_3}&=&\left (\frac{16}{9}g_3^2 - 
\frac{m_{top}^2}{v_{top}^2}\right )m_{top}\label{e7a}\\
g_3^3\frac{dv_{top}}{dg_3}&=&\left (\frac{m_{top}^2}{v_{top}^2}\right )
v_{top}\label{e7b}
\end{eqnarray}

Following reference~\cite{r3}, we introduce $\rho_{top}(g_3)=\frac{m_{top}^2}{g_3^2}$. 
From equation (\ref{e7a}), this satisfies the equation
\begin{equation}
g_3 \frac{\rho_{top}^2}{g_3}=\frac{14}{9}\rho_{top}- 
2\frac{\rho_{top}^2}{v_{top}^2}\label{e7c}
\end{equation}
To simplify further, we set $\rho_{top}=\Gamma_{top}v_{top}^2$ and using equation
(\ref{e7b}) and (\ref{e7c}), we get
\begin{eqnarray}
g_3\frac{d(\Gamma_{top}v_{top})}{ dg_3 }&=&g_3\left[ \frac{ d\Gamma_{top} }{ dg_3 }
v_{top}^2 +2\Gamma_{top}v_{top} \frac{ dv_{top} }{ dg_3 }\right ]\\
&=&g_3\left[ \frac{d\Gamma_{top}}{dg_3}v_{top}^2
+2\Gamma_{top}\frac{m_{top}^2}{
v_{top}^2g_3^3}\right ]\\
&=&g_3v_{top}^2 \frac{d\Gamma_{top}}{dg_3}v_{top}^2+2\Gamma_{top}v_{top}^2\label{e7d}
\end{eqnarray}
Equating equation (\ref{e7c}) to equation (\ref{e7d}), we obtain
\begin{equation}
g_3\frac{d\Gamma_{top}}{dg_3}=\frac{14}{9}\Gamma_{top}-4\Gamma^2_{top}
\end{equation}
The equation, exhibiting the pole, is
\begin{equation}
\frac{d\Gamma_{top}}{\Gamma_{top}\left ( \frac{7}{18}-\Gamma_{top}\right )}
=4\frac{dg_3}{g_3},\label{e14a}
\end{equation}
On integration,
\begin{equation}
\Gamma_{top}=\frac{7}{18}\left [ \frac{g_3^{\frac{14}{9}}}
{C_{top}+g_3^{\frac{14}{9}}}\right ],\label{e15}
\end{equation}
$C_{top}$ is independent of $g_3$. This is similar to or the  general solution of
Kubo et al~\cite{r3} for $g_1=g_2=$0.
$C_{top}$ may depend on the other gauge couplings $g_1$,$g_2$
at higher energies without invalidating equation (\ref{e15}). We solve the 
equation for the $v_{top}$ by noting that 
\begin{equation}
\frac{m_{top}^2}{g_3^2v_{top}^2}= \frac{\rho_{top}}{v_{top}^2}=\Gamma_{top}
\end{equation}
On simplification
\begin{equation}
\frac{dv_{top}}{v_{top}}=\frac{7}{18}\frac{ g_3^{ \frac{14}{9} } }
{C_{top}+g_3^{\frac{14}{9}}}\frac{dg_3}{g_3}.\label{e15a}
\end{equation}
This can be deduced to be the same as the corresponding equation for $m_{top}$
of Faraggi~\cite{r3a} translated to our notation. Integrating equation (\ref{e15a}),
\begin{equation}
v_{top}^4=v^4(C_{top}+g_3^{ \frac{14}{9} })
\end{equation}
Using $ v_{top}^2$ in  $m_{top}^2=M_{top}^2v_{top}^2$, one obtains the solutions of 
(\ref{e7a}) and (\ref{e7b})
\begin{eqnarray}
m_{top}&=&\sqrt{\frac{7}{18}}g_3^{-\frac{7}{9}} g_3\left[ \frac{g_3^{\frac{14}{9}} }
{\left ( C_{top}+g_3^{\frac{14}{9}}\right )^{\frac{1}{4}}} \right ]v,\label{e8a}\\
 v_{top}&=&v(C_{top}+g_3^{ \frac{14}{9}})^{\frac{1}{4}}\label{eQ}
\end{eqnarray}
This can be confirmed by direct substitution.

Observationally, one intends to know the value of $m_{top}(t)$ for a given mass $m_1$, 
such that $t_1=log(\frac{m_1}{M_Z})$.
The function in the square bracket, optimally stable like the equation (\ref{e15}),
with respect to the variation of $g_3$ at this mass $M_1(t_1)$ or t=$t_1$, should be
\begin{equation}
\frac{\partial}{\partial g_3} \frac{g_3^{\frac{14}{9}} }
{\left ( C_{top}+g_3^{\frac{14}{9}}\right )^{\frac{1}{4}}} =0
\end{equation}
or
\begin{equation}
C_{top}(t_1)=-\frac{3}{4}g_3^{\frac{14}{9}}(t_1).\label{e7e}
\end{equation}
Finally, we get from (\ref{e8a}), (\ref{eQ}) and (\ref{e7e}), the solutions of coupled equations (\ref{e7a}) and (\ref{e7b}) as
\begin{eqnarray}
m_{top}(t)&=&\sqrt{\frac{7}{9}}~ g_3^{\frac{7}{9}}(t)~ g_3(t)   
\left (\frac{1}{ 4g_3^{\frac{14}{9}}(t)-3g_3^{\frac{14}{9}}(t_1) }\right )^{\frac{1}{4}}v
\label{e9b}\\
v_{top}(t)&=&\sqrt{\frac{1}{2}} \left( 4g_3^{\frac{14}{9}}(t)
-3g_3^{\frac{14}{9}}(t_1) \right )^{\frac{1}{4}}v\label{e9a}\\
M_{top}^2(t)&=&\frac{m_{top}^2}{v_{top}^2}=\frac{14}{9} g_3^{\frac{14}{9}}(t) g_3^2(t)
\frac{1}{\left( 4g_3^{\frac{14}{9}}(t)-3g_3^{\frac{14}{9}}(t_1) \right )}\label{e9e}
\end{eqnarray}
The 'stable' values of $m_{top}$ and $v_{top}$, at $t_1$, based 
on the equation (\ref{e7e}), are
\begin{eqnarray}
m_{top}(t_1)&=&\sqrt{\frac{7}{9}}g_3^{\frac{7}{18}}(t_1)g_3(t_1)~v\label{e9d}\\
Sin(\beta(t_1))&=&\sqrt{\frac{1}{2}}g_3^{\frac{7}{18}}(t_1)\label{e9c}
\end{eqnarray}
for comparision with experimental results at mass for which t=$t_1$.

We report the following results
\begin{eqnarray}
m_{top}(m_Z)&=&180.92 ~GeV\\
m_{top}(m_X)&=&m_U=110 ~GeV\label{e9f}\\
tan\beta(M_Z)&=&1.1\\
tan\beta(m_X)&=&0.9
\end{eqnarray}

These are the results from the explicitly known  asymptotically free region.
Experimentally, the top quark has a
mass of 174.3$\pm$ 5.1\cite{pdg} GeV and in excellent agreement with our result. 
Eventhough the extrapolation is too drastic, the decrease in the  value of 
running tan$\beta$ as the mass changes fron $M_Z$ to $M_X$, confirms the result 
which was first reported by Parida and Purkayastha~\cite{r4}. The value of tangent 
hovers around one. So, to get a overall picture of the general nature of solutions 
of the MSSM R.G. Equations, it is fairly adequate to solve the 
equations for the fermion couplings $M_F$ like $M_{top}$ of equation(\ref{e3}).

We continue to expand the parameter space, retaining the top and bottom Yukawa
couplings(M$_{top}$,M$_{bottom}$) and color gauge coupling g$_3$. The equations are
~\cite{r3a}
\begin{eqnarray}
16\pi^2\frac{d~}{dt}log~M_{top}&=&\left [-\frac{16}{3}g_3^2 +6M_{top}^2
+M_{bottom}^2\right ]\label{e16aa}\\
16\pi^2\frac{d~}{dt}~log~ M_{bottom}&=&\left [-\frac{16}{3}g_3^2 +M_{top}^2
+6M_{bottom}^2\right ]\label{e16ab}
\end{eqnarray}
We eliminate M$_{bottom}$ from r.h.s, by noting
\begin{equation}
16\pi^2\frac{d~}{dt}~ log(\frac{M_{top}^6~~~}{M_{bottom}})=
\left [-\frac{80}{3}~g_3^2 +35~M_{top}^2\right ]\label{e16ac}
\end{equation}
From equation (\ref{e8a})
\begin{equation}
M^2_{top}=\frac{m_{top}^2}{v_t^2~}=\frac{7}{18}~g_3^2\frac{g_3^{\frac{14}{9}}}
{C_{top}+g_3^{\frac{14}{9}}}
\end{equation}
This is easily solved for $M_{bottom}$. After some algebra, we get at t=$t_1$=0
i.e. $M_Z$
\begin{equation}
M_{bottom}(M_Z)=2^{\frac{1}{6}}(\frac{7}{18})^3~g_3^{-\frac{103}{108}}
\end{equation}
hence
\begin{equation}
m_{bottom}=M_{bottom}v_{bottom}=M_{bottom}(v^2-v^2_{top})^{\frac{1}{2}}\cong 5.31~ GeV
\end{equation}
This is a very good result in view of the shrunk region of parameter space.

As a step further, we expand the region of validity of 
perturbation to include SU(2) and U(1) i.e. $g_2\neq 0$ and $g_1\neq 0$. Then
the equation for top becomes
\begin{equation}
16\pi^2\frac{dM_{top}}{dt~~~}=\left [- \frac{13}{15}g_1^2 - 3g_2^2-\frac{16}{3}g_3^2
+6M_{top}^2\right ]M_{top}\label{e17}
\end{equation}
This equation (\ref{e17}) has been exactly solved by Deo and Maharana~\cite{r5} and
the result is
\begin{equation}
1=\frac{M_{top}^2} {m_U^2} ~a_{top} + b_{top}\label{e18}
\end{equation}
where
\begin{eqnarray}
 a_{top}&=&\left ( \frac{g_1^2}{g_U^2}\right )^{\frac{K_1^U}{c_1}}
\left ( \frac{g_2^2}{g_U^2}\right )^{\frac{K_2^U}{c_2}}
\left ( \frac{g_3^2}{g_U^2}\right )^{\frac{K_3^U}{c_3}},\\
\text{and}~~~~ b_{top}&=&\frac{6}{8\pi^2} \int _0^{t_X} dt~ 
\left (1- g_1^2\frac{c_1}{8\pi^2} t\right )^{\frac{K_1^U}{c_1}}
\left (1- g_2^2\frac{c_2}{8\pi^2} t\right )^{\frac{K_2^U}{c_2}}   
\left (1- g_3^2\frac{c_3}{8\pi^2} t\right )^{\frac{K_3^U}{c_3}}.
\end{eqnarray}
$K_1^U=13/15$, $ K_2^U=3$ and $K_3^U=16/3$ are the coefficients 
of the coupling constant as given in references~\cite{r5,r6}.
We found that the top mass originated from the unification mass $m_U\sim$ 114 GeV.

This also confirms the perturbative `stability' of this approach. 
Taking $g_1\neq $0 and $g_2\neq $0, and extending to large t-values,
the value of $m_U$ of equation (\ref{e9f}), changes by a few GeV only.

In reference~\cite{r5}, it has been shown that all the 12 fermions
(quarks and leptons) at the GUT scale had the same mass of about 115 GeV. 
Encouraged by this, we now enlarge the region of applicability of the analysis for 
the whole parameter space. Essentially, there are 13 parameters, 12 fermion 
masses and precisely one CKM phase angle $\phi$ for the three generations. 
We now write the full renormalization equation for all the fermions,
\begin{eqnarray}
16\pi^2\frac{dM_F(t)}{dt~~~~}&=&A_FM_F^3(t)+[Y_F(t)-G_F(t)]M_F(t)\label{e19}\\
&=&A_FM_F^3(t)+Z_F(t)M_F(t).\label{e20}
\end{eqnarray}
$A_F$ is a group theoretic factor whose value is `6' for quarks, i.e. for
$F$=1,2,$\cdots$,6 and `4' for the leptons i.e. for $F$=7,8,$\cdots$ , 12. 
The positive values indicate that the field theory containing Yukawa 
couplings only, may not be asymptotically free.

$Y_F$ is the mixing term which can be put in matrix form
\begin{equation}
Y_F=\sum_H A_{FH}M_H^\dagger(t) M_H(t),~~~~~~~H=1,2, \cdots , 12.\label{e21}
\end{equation}
In MSSM, the matrix $A_{FH}$ is specified by the 144 elements given below~\cite{r6,r7},
\begin{eqnarray}
A_{FH}=
\left (
\begin{array}{cccccccccccc}
 0 & 3 & 3 & 1 & 0 & 0 & 0 & 0 & 0 & 1 & 1 & 1\\
3 & 0 & 3 & 0 & 1 & 0 & 0 & 0 & 0 & 1 & 1 & 1\\
3 & 3 & 0 & 0 & 0 & 1 & 0 & 0 & 0 & 1 & 1 & 1\\
1 & 0 & 0 & 0 & 3 & 3 & 1 & 1 & 1 & 0 & 0 & 0\\
0 & 1 & 0 & 3 & 0 & 3 & 1 & 1 & 1 & 0 & 0 & 0\\
0 & 0 & 1 & 3 & 3 & 0 & 1 & 1 & 1 & 0 & 0 & 0\\
0 & 0 & 0 & 3 & 3 & 3 & 0 & 1 & 1 & 1 & 0 & 0\\
0 & 0 & 0 & 3 & 3 & 3 & 1 & 0 & 1 & 0 & 1 & 0\\
0 & 0 & 0 & 3 & 3 & 3 & 1 & 1 & 0 & 0 & 0 & 1\\
3 & 3 & 3 & 0 & 0 & 0 & 1 & 0 & 0 & 0 & 1 & 1\\
3 & 3 & 3 & 0 & 0 & 0 & 0 & 1 & 0 & 1 & 0 & 1\\
3 & 3 & 3 & 0 & 0 & 0 & 0 & 0 & 1 & 1 & 1 & 0
\end{array}
\right ),\label{e22}
\end{eqnarray}
Then one can write an exact solution for $M_F(M_Z)$, as given in reference~\cite{r7}
\begin{equation}
\frac{M_{top}^2(M_Z)}{M_F^2(M_Z)}=\frac{M_{top}^2(M_Z)}{M_F^2(M_X)}
\exp \left( \frac{1}{8\pi^2}\int_0^{t_X}Z_F(\tau) d\tau \right)
+\frac{A_F}{8\pi^2}\int_0^{t_X}dt~
\exp \left( \frac{1}{8\pi^2}\int_0^{t} Z_F(\tau) d\tau\right)\label{e25}
\end{equation}
The first term of equation (\ref{e20}) is $A_F~M_F^3$ which is 
perturbatively very small except for the top. Therefore we neglect 
the 2nd term in (\ref{e25}). For fermions other than top, 
i.e. F varying from 2 to 12,
\begin{eqnarray}
M_F(M_Z)&\simeq & m_U~e^{-\frac{1}{16\pi^2}\int_0^{t_X}Z_F(\tau)~d\tau},\\
&\simeq & m_U~e^{-\frac{I_F}{16\pi^2}}, \label{e26}
\end{eqnarray}
where 
\begin{equation}
I_F=\int_0^{t_X}Z_F(t)~dt.
\end{equation}
But, the main problem of finding general solutions, is to calculate 
$I_F$. This integral is
\begin{equation}
I_F=\int_0^{t_X}Z_F(t)dt=\frac{1}{2} \int_0^{t_X}[Z_F(t)+Z_F(-t)]dt
=~\frac{1}{4}\int_{-t_X}^{t_X}dt[Z_F(t)+Z_F(-t)] \label{e27}
\end{equation}

Next, we consider the equality,
\begin{equation}
dt=\frac{dt}{dM_F} dM_F(t)=16\pi^2 \frac{dM_F(t)}{M_F\left ( 
A_F M_F^{\dagger}M_F + \sum_H A_H M_H^{\dagger}M_H-G_F(t)
\right )}\label{e28}
\end{equation}
The integrand should have the poles like the ones in equation (\ref{e14a}). 
The most solutions have $M_{lepton}$=0 as the gauge factor $G_{lepton}$ 
does not contain the color gauge coupling g$_3$. So,we shall let 'H'
to be summed over from 1 to 6 in $\sum_HA_{FH}$. Choosing 
\begin{equation}
M_F(t)=m_Ue^{in_F\theta_F(t)},
\end{equation}
we have 
\begin{eqnarray}
I_F&=&i\frac{1}{2}~16\pi^2n_F\int_{-m_U}^{m_U}d\theta_F(t)
\frac{ \sum_H A_{FH} -\frac{1}{m_U^2}G_F(\theta(t))}{
A_F + \sum_H A_{FH} - \frac{1}{m_U^2}G_F(\theta(t))}\\
&\cong&i\frac{1}{2}~16\pi^2n_F\frac{1}{12}\sum_G\sum_HA_{GH}
\int_{-t_X}^{t_X}d\theta_G(t)\frac{1}{A_F + \sum_H A_{FH}- 
\frac{1}{m_U^2}G_F(\theta(t))}
\end{eqnarray}
where $n_F$ in an integer. We have neglected $\frac{1}{m_U^2}G_F(t)$ 
in the numerator and want to make the factor 
multiplying $n_F$, independent of F by taking the average, using 
$\frac{1}{12}\sum_G$ =1. Retracing back, we replace 
$M_F(t)=m_Ue^{i\theta_F(t)m_U^2}$  and find that
\begin{equation}
16\pi^2\frac{dM_G(t)}{M_G(t)}=id\theta_G(t)m_U^2
=(A_G+\sum_{H=1}^6A_{GH}  - \frac{1}{m_U^2}G_G(\theta(t)))m_U^2
\end{equation}
so that 
\begin{equation}
I_F=n_Ft_X\frac{1}{12}\sum_{G=2}^{12}\sum_{H=1}^6A_{GH}=n_F t_X\frac{89}{12}
\end{equation}
So the masses
of the fermions other than the top is
 
\begin{equation}
M_F(M_Z)\simeq m_U \lambda^{n_F}\label{e35}
\end{equation}

The Wolfenstein parameter $\lambda $ turns out to be
\begin{equation}
\lambda=\exp\left(-\frac{t_X}{16\pi^2} \frac{89}{12}\right)=0.219\label{e34}
\end{equation}
This is an excellent result in spite of the approximate estimates.

Starting from the unification mass $m_U$ =115 GeV and using the 
equation (\ref{e35}), we now calculate masses of all 11 fermions for 
different values of $n_F$ and identify them in the Table-\ref{tab:table1}. 
\begin{center}
\begin{table}[h]
\begin{tabular}{||c|c|c|c||}\hline\hline
$n_F$&Mass  & Fermions&Expt. Values\\
    & (GeV)   &(Quarks, leptons)&(GeV)\\         
\hline
2 &5.5&b&5.\\ \hline
3 &1.2& c,~$\tau$ & 1.4, ~1.7\\ \hline
4 &0.264&s,~ $\mu$ & 0.1~-~0.23,~ 0.01\\ \hline
5 & 0.06& ? &?\\ \hline
6 &0.012&d,~$\nu_{\tau}$ & 0.055~-~0.115,~ 0.018\\ \hline
7 &0.0028&u&0.003 \\ \hline
8 & 6$\times 10^{-4}$ & e&5$\times 10^{-4}$ \\ \hline
9 & 1.33 $\times 10^{-4}$ & $\nu_{\mu}$&1.9$\times 10^{-4}$\\ \hline
16& 3$\times 10^{-9}$& $\nu_e$&3$\times 10^{-9}$\\\hline\hline
\end{tabular}
\caption{\label{tab:table1}Identification of fermions}
\end{table}
\end{center}
The values given in the table-\ref{tab:table1} are estimates only based, on the relation 
(\ref{e35}). The remaining free parameter is the CKM phase angle $\phi$ 
defined through Wolfenstein parametrization~\cite{r8} and is given as~\cite{r8a}
\begin{equation}
\lambda= \left ( \frac{M_d}{M_s} +\frac{M_u}{M_c} +
2\sqrt{\frac{M_d}{M_s}\frac{M_u}{M_c}}\cos\phi\right )^{\frac{1}{2}}.\label{e36}
\end{equation}
Using the Table-\ref{tab:table1} for the values of $n_F$, we get
\begin{equation}
cos\phi\simeq -\frac{\lambda}{2}\simeq -0.1\label{e37}
\end{equation}
and this gives $\phi \simeq 95^o$.

The mixing angles $c_i,s_i$, i=1,2,3,4 of the CKM matrix as given for quarks is
\begin{eqnarray}
V_{CKM}=
\left (
\begin{array}{ccc}
c_1c_2-s_1s_2e^{-i\phi}& s_1+c_1s_2e^{-i\phi}&s_2(s_3-s_4)\\
-c_1s_2-s_1e^{-i\phi}&-s_1s_2+(c_1c_2c_3c_4+s_1s_4)e^{-i\phi}&s_3-s_4\\
s_1(s_3-s_4)&-c_1(s_3-s_4)& (c_3c_4+s_1s_4)e^{i\phi}\\
\end{array}\label{eq69a}
\right )
\end{eqnarray}
From table-I,
\begin{equation}
s_1=\left (\frac{M_d}{M_s}\right )^{\frac{1}{2}}=\lambda,~~~
s_2=\left (\frac{M_u}{M_c}\right )^{\frac{1}{2}}=\lambda^2;~~~~
s_4=\left (\frac{M_dM_s}{M_b^2}\right )^{\frac{1}{2}}=\lambda^3~~\text{and}~~
s_3-s_4=\lambda^2A(t)
\end{equation}
The SUSY RG equations are~\cite{r9},
\begin{equation}
\frac{d\lambda}{dt}=0,
\end{equation}
and
\begin{equation}
16\pi^2\frac{d}{dt} logA(t)=-(M_{top}^2+M_b^2).\label{e54}
\end{equation}
So $s_1, s_2,s_4,$ and $\phi$ do not change with energy. Further more, to 
satisfy the equation (\ref{e54}), we must equate A(t)  as 
\begin{equation}
A(t)=\left [ \frac{M_b(t)M_{top}(t)}{M_{top}^2(M_Z)}\right ]^{-1/7},
~~~~~~A(M_X)=1.12~~~~~\text{and}~~~~~ A(M_Z)=1.5 .\label{eq85}
\end{equation}
The entire CKM matrix elements are calculable. Choosing $A(M_X)$=1.1 in the CKM matrix
for all t, we can determine $m_U$ from $M_{top}$ and then the rest of the masses
from the equation (\ref{e35}) and, as in reference~\cite{r7},
\begin{equation}
M_F(t)=m_U \lambda^{n_F \left(1-\frac{t}{33}\right)}
\end{equation}

Thus we are able to find all the free parameters
in the matter sector of MSSM, subject to the accuracy of the above relation.
In the gauge sector, the two parameters $M_X$ and $\alpha_{GUT}$ are the 
only inputs in the model.

In this letter, we have found the top mass from the MSSM RGE, using the group 
theoretic coefficients, as 180 GeV in close agreement with experiment.
Using group constants, we have calculated a unification mass and have been
able to find an approximate equation for the eleven  other fermions, the CKM phase angle 
$\phi$ in terms of the Wolfenstein parameter. The approach made above to solve RGE
should be studied further with greater detail for the eleven fermions separately and
much more accurately.

\end{document}